\documentclass[journal]{IEEEtran}

\ifCLASSINFOpdf
\else
   \usepackage[dvips]{graphicx}
\fi
\usepackage{url}

\usepackage{microtype}
\usepackage{graphicx}
\usepackage{svg}
\usepackage{amsmath}
\usepackage{cite}

\usepackage{enumitem}

\DeclareMathOperator*{\argmax}{arg\,max}

\usepackage{booktabs}
\usepackage{tabularx}
\usepackage{multirow}
\newcommand{\mytable}{
	\centering
	\renewcommand{\arraystretch}{1.1}
}

\newcolumntype{C}{>{\centering\arraybackslash}X}
\newcolumntype{L}{>{\raggedright\arraybackslash}X}
\newcolumntype{R}{>{\raggedleft\arraybackslash}X}
\newcolumntype{P}[1]{>{\raggedright\arraybackslash}p{#1}}

\newcommand{\PreserveBackslash}[1]{\let\temp=\\#1\let\\=\temp}
\newcolumntype{A}[1]{>{\PreserveBackslash\raggedright}p{#1}}
\newcolumntype{B}[1]{>{\PreserveBackslash\centering}p{#1}}

\newcommand{\figcaptionsep}{\vspace*{-7.5pt}}
\newcommand{\tablecaptionsep}{\vspace*{-5pt}}

\begin{document}

\title{Rhythm Modeling for Voice Conversion}

\author{Benjamin van Niekerk, Marc-André Carbonneau, Herman Kamper
\thanks{B.~van Niekerk and H.~Kamper are with the Department of Electrical and Electronic Engineering, Stellenbosch University, South Africa (e-mails: benjamin.l.van.niekerk@gmail.com and kamperh@sun.ac.za).} 
\thanks{M.-A.~Carbonneau is with Ubisoft La Forge, Montréal (e-mail: macarbonneau@ubisoft.com).}}

\markboth{Submitted July 2023}
{Shell \MakeLowercase{\textit{et al.}}: Bare Demo of IEEEtran.cls for IEEE Journals}
\maketitle

\begin{abstract}

Voice conversion aims to transform source speech into a different target voice. 
However, typical voice conversion systems do not account for rhythm, which is an important factor in the perception of speaker identity.
To bridge this gap, we introduce Urhythmic---an unsupervised method for rhythm conversion that does not require parallel data or text transcriptions. 
Using self-supervised representations, we first divide source audio into segments  approximating sonorants, obstruents, and silences. 
Then we model rhythm by estimating speaking rate or the duration distribution of each segment type.
Finally, we match the target speaking rate or rhythm by time-stretching the speech segments.
Experiments show that Urhythmic outperforms existing unsupervised methods in terms of quality and prosody.

\end{abstract}

\begin{IEEEkeywords}
voice conversion, rhythm conversion, speaking rate estimation
\end{IEEEkeywords}

\IEEEpeerreviewmaketitle

\section{Introduction}

\IEEEPARstart{F}{rom} a slow, purposeful oration to rapid, excitable chatter, rhythm conveys emotion and intent in speech.
Rhythm and speaking rate are also important cues for identifying different speakers \cite{van1987contribution}.
Despite its role in communication, typical voice conversion systems do not model the target speaker’s rhythm, reproducing the prosody of the source speech instead.

\begin{figure}[t]
\centerline{\includegraphics[width=0.95\columnwidth]{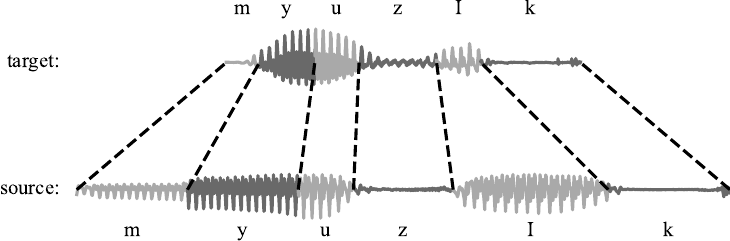}}
\figcaptionsep
\caption{
    An example of the rhythm conversion problem.
    Both  utterances contain the word `music', but differ in rhythm.
    The source utterance is spoken at a slower rate than the target.
    However, the average speaking rate does not capture fine-grained differences in rhythm.
    For example, the sonorants [m], [y], [u], and [I] are shortened far more than
    the obstruents [z] and [k].
}
\label{fig:rhythm}
\end{figure}

Consider the pair of utterances shown in Figure~\ref{fig:rhythm}. 
Both contain the word but differ in rhythm. 
Most noticeable is the speaking rate---the source utterance is spoken more slowly than the target.
Besides speaking rate, features like prolonged vowels or brief pauses characterize individual rhythm and are influenced by accent~\cite{deterding2001measurement, torgersen2011study}, gender~\cite{wassink2001theme}, or even historical period~\cite{thomas2006prosodic}.
Our goal is to better convert speaker identity by modeling the natural rhythm of the target speaker. 

Some recent work explores rhythm conversion using sequence-to-sequence models \cite{zhang2019sequence, tanaka2019atts2s} or forced alignment~\cite{csicsman2017transformation}.
However, training these systems requires parallel speech or text transcriptions, which are costly and time-consuming to collect.
Unsupervised methods such as AutoPST~\cite{qian2021global}, UnsupSeg~\cite{kuhlmann2022investigation}, and DISSC~\cite{maimon2022speaking} lift this restriction by modeling rhythm without annotations or parallel data.
However, there is still a gap in quality and prosody compared to natural speech. 
For example, AutoPST and DISSC discard some content information, resulting in poor intelligibility.
UnsupSeg only models rhythm globally---ignoring fine-grained details.

To tackle these problems, we propose Urhythmic\footnote{Code and checkpoints: \url{https://github.com/bshall/urhythmic}}\footnote{Audio samples: \url{https://ubisoft-laforge.github.io/speech/urhythmic/}}, an unsupervised approach to rhythm conversion and control. 
Building on self-supervised speech representations~\cite{mohamed2022self}, Urhythmic models both global and fine-grained characteristics of rhythm.
As the foundation of Urhythmic, we divide source audio into discovered, variable-duration segments approximating sonorants (vowels, approximants, and nasals), obstruents (fricatives and stops), and silences.
Based on this segmentation, we estimate speaking rate by counting the number of sonorants per second of speech.
For fine-grained modeling, we approximate the duration distribution of each segment type.
Finally, we time-stretch the entire utterance or individual segments to match the target speaking rate or rhythm.

Our main contributions are:
\begin{enumerate}
    \item We propose Urhythmic, a voice and rhythm conversion system that does not require text or parallel data.
    \item We develop global and fine-grained methods based on speaking rate estimation or segment duration modeling.
    \item We show that both methods outperform existing approaches.
    However, fine-grained modeling more effectively matches the target speaker's pattern of pauses and silences.
\end{enumerate}

\section{Proposed Method}

A simple method for rhythm modeling is to use time-aligned transcriptions to estimate speaking rate. 
Alternatively, we can capture finer-grained characteristics by modeling the duration distribution of individual phones or syllables.
Then, we can alter rhythm by time-stretching a source utterance to match the target speaking rate or duration distributions. 
However, this approach requires text transcriptions and forced alignment, which are not available in many voice conversion applications.

\begin{figure}[t]
\centerline{\includegraphics[width=0.99\columnwidth]{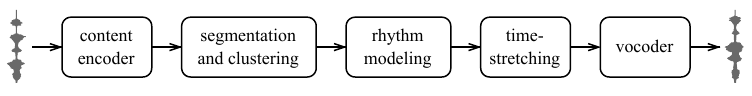}}
\vspace*{-10pt}
\caption{
    Overview of Urhythmic, our unsupervised rhythm conversion model.
}
\label{fig:overview}
\end{figure}

To remove the need for transcriptions, we segment
speech into sonorants, obstruents, and silences without supervision.
We model the duration of these segments to characterize rhythm.
Fig.~\ref{fig:overview} shows an overview of our approach.
First, the content encoder translates input audio into speech units (Sec.~\ref{sec:encoding}).
Next, the segmentation and clustering block groups similar units into short segments. 
The segments are then combined into coarser groups corresponding to sonorants, obstruents, and silences (Sec.~\ref{sec:segmenting}). 
The rhythm modeling block estimates speaking rate or models the duration distribution of each group (Sec.~\ref{sec:rhythm-modeling}). 
The time-stretching block down/up-samples the speech units to match the target rhythm or speaking rate (Sec.~\ref{sec:time-stretching}). 
Finally, the vocoder converts the speech units into an audio waveform.

\subsection{Content Encoding}
\label{sec:encoding}

The content encoder aims to extract speech representations that capture linguistic content but discard speaker-specific details.
By replacing the speaker information, we can convert source speech to a target voice.
To achieve this, we encode source audio into a sequence of soft speech units $\langle \mathbf{s}_1, \ldots, \mathbf{s}_T \rangle$. 
Soft units were proposed as an alternative to discrete speech representations for voice conversion \cite{van2022comparison}.
While discretization acts as a bottleneck to remove speaker information \cite{van2017neural, van2020vector, polyak2021speech}, it also discards some linguistic content---increasing mispronunciations in converted speech.
To avoid this problem, \cite{van2022comparison} trains a soft content encoder to predict a distribution over discrete units.
By modeling uncertainty, the soft units retain more content information and improve voice conversion. 
Concretely, each soft unit parameterizes a distribution over a dictionary of discrete speech units:
\[
p(i \mid \mathbf{s}_t) = \frac{\exp(\text{sim}(\mathbf{s}_t, \mathbf{e}_i) / \tau)}{\sum_{k=1}^K \exp(\text{sim}(\mathbf{s}_t, \mathbf{e}_k) / \tau)},
\]
where $i$ is the index of the $i^\textrm{th}$ discrete unit, $\mathbf{e}_i$ is a corresponding embedding vector, and $\text{sim}(\cdot, \cdot)$ computes the cosine similarity between the soft and discrete units.

\subsection{Segmentation and Clustering}
\label{sec:segmenting}

The segmentation and clustering block groups speech into variable-duration segments. 
First, we partition the soft units into short segments based on \cite{kamper2020towards}. 
We frame segmentation as an optimization problem, maximizing the similarity between the units within a segment. 
Next, using hierarchical clustering, we merge the segments into larger groups approximating sonorants (vowels, approximants, and nasals), obstruents (fricatives and stops), and silences.

In the first step, we partition the soft units into a sequence of contiguous segments $\langle g_1, \ldots, g_N \rangle$.
Each segment $g_n = (a_n, b_n, i_n)$ is defined by a start index $a_n$, an end index $b_n$, and a single representative discrete unit $i_n$.
We assess the quality of a given segmentation by scoring it according to the soft unit predictor:
\[
\mathcal{E}(\mathbf{s}_{1:T}, g_{1:N}) = \sum_{g_n \in g_{1:N}} \sum_{t = a_n}^{b_n} \log p(i_n \mid \mathbf{s}_t) + \gamma (b_n - a_n).
\]
Here, a higher score corresponds to a better segmentation.
The last term in the summation is a regularizer encouraging longer segments, with $\gamma$ controlling the importance of the term.
Note that without the regularizer, the optimal segmentation places each soft unit in its own segment.

\begin{figure}[!b]
\centerline{\includegraphics[width=\columnwidth]{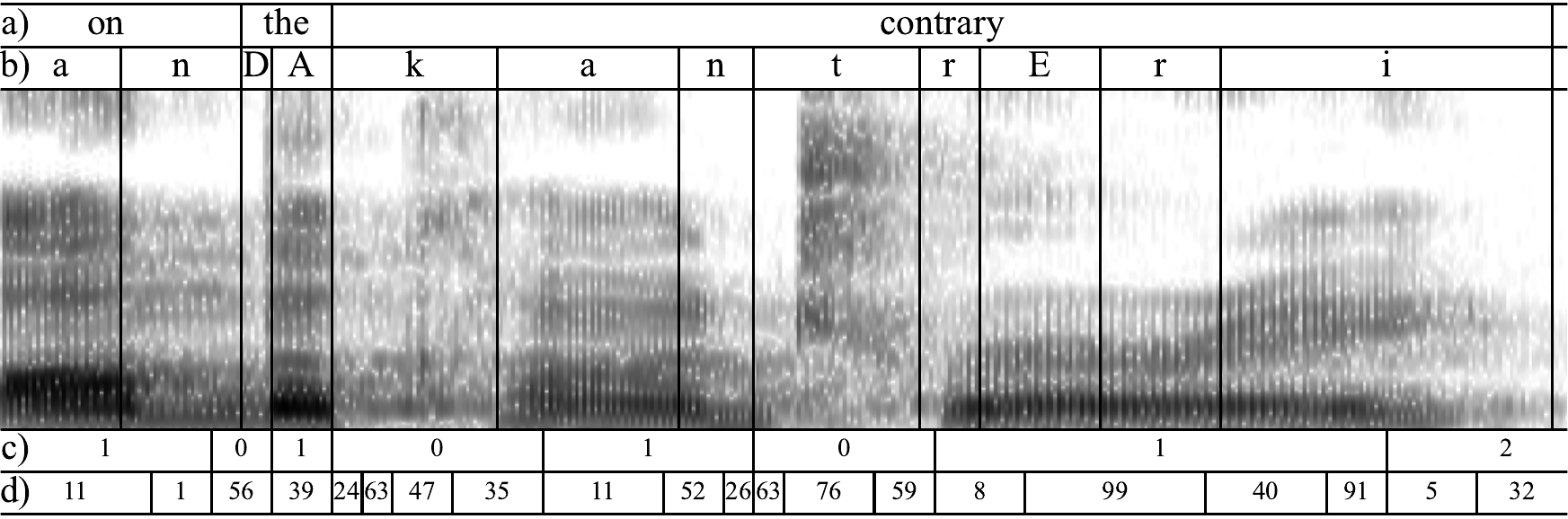}}
\figcaptionsep
\caption{
    An example segmentation of the utterance `on the contrary'.
    a) Aligned word boundaries.
    b) Aligned phonetic transcription.
    c) Hierarchical grouping of the segments into three clusters corresponding to sonorants, obstruents, and silences. 
    d) Segmentation of the soft speech units.
}
\label{fig:spectrogram}
\end{figure}

We can find the best segmentation: 
\[
\mathrm{g}^\star_T = \argmax_{g_{1:N}} \mathcal{E}(\mathbf{s}_{1:T}, g_{1:N}),
\] 
by applying dynamic programming to the recurrence:
\[
\mathcal{E}(\mathbf{s}_{1:T}, \mathrm{g}^\star_T) = \min_{t<T} \left( \mathcal{E}(\mathbf{s}_{1:t}, \mathrm{g}^\star_t) + \min_{g} \mathcal{E}\left(\mathbf{s}_{t+1:T}, \langle g \rangle \right) \right).
\]
Fig.~\ref{fig:spectrogram} row (d) shows an example segmentation partitioning the utterance into short sub-phone units. 
In preliminary experiments, we found that these shorter segments are not ideal for rhythm modeling. 
So to combine segments into larger groups, we hierarchically cluster the dictionary of discrete units and merge adjacent segments belonging to the same cluster. 

\begin{figure}[t]
\centerline{\includegraphics[width=0.9\columnwidth]{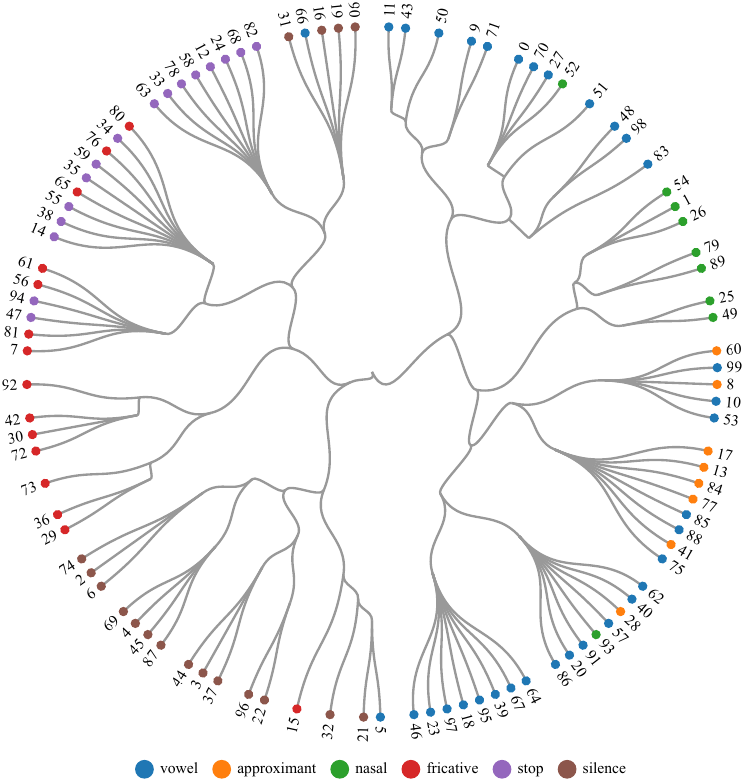}}
\vspace*{-5pt}
\caption{
    Dendogram showing the clusters of discrete speech units.
    Each unit is colored by the most frequently overlapping sound type.
    The three main branches correspond to sonorants (vowels, approximants, and nasals), obstruents (fricatives and stops), and silences.
}
\label{fig:dendogram}
\end{figure}

Fig.~\ref{fig:dendogram} visualizes the dendrogram constructed through agglomerative clustering. 
We label each segment according to the three main branches of the dendrogram.
Then we join adjacent segments with the same label. 
Fig.~\ref{fig:spectrogram} row~(c) shows the merged segments. 
Referring to the phonetic transcription in row~(b), we see that the larger segments approximate sonorants, obstruents, and silences.
To validate this observation, we color each discrete unit in Fig.~\ref{fig:dendogram} by the most frequently overlapping sound type (vowel, approximant, nasal, fricative, stop, or silence). 
The dendrogram clearly clusters the units by sound type, with the three main branches representing sonorants, obstruents, and silences.
Note that phonetic transcriptions are only used for analysis and are not required by our approach. 

\subsection{Rhythm Modeling}
\label{sec:rhythm-modeling}

Building on the segmentation block, we propose two methods for rhythm modeling. 
The first method estimates the global speaking rate. 
The second models the duration of individual segments, providing a finer-grained characterization of rhythm.

Speaking rate is typically measured in syllables per second
\cite{grosjean1976listener, pfau1998estimating}. 
With time-aligned transcriptions, we can calculate the syllable rate by simply counting the number of syllables and dividing by the total duration.
Without transcriptions, we count sonorant segments as an approximation.
For example, row~(c) of Fig.~\ref{fig:spectrogram} has four sonorant segments (labeled~1) over a duration of 0.94 seconds, giving an estimated rate of 4.26 segments per second.
Since sonorants generally correspond to syllable nuclei~\cite{anderson2018essentials}, this approximation should correlate with the syllable rate.
We measure the correlation in Sec.~\ref{sec:exp-speaking-rate}.

To estimate the speaking rate, we need to identify which clusters correspond to sonorants, obstruents, and silences.
We classify the clusters based on energy and voicing features. 
First, we mark silent intervals using an energy-based voice activity detector. 
Then for each cluster, we calculate the percentage overlap with the marked intervals. 
The cluster with the highest overlap is labeled as silence.
To distinguish between sonorants and obstruents, we apply a similar method to voicing flags extracted by a pitch detector.
Specifically, we label the cluster that most frequently overlaps with voiced speech as sonorant.

Our second approach aims to model finer-grained rhythm information. 
Instead of estimating the global speaking rate, we model the duration distribution of each cluster to capture variations in pauses, vowel length, etc.
Historically, duration modeling has been an important component in text-to-speech \cite{van1994assignment} and speech recognition systems \cite{levinson1986continuously, johnson2005capacity}.
In particular, \cite{levinson1986continuously} and \cite{johnson2005capacity} apply parametric models to phone durations, showing that the gamma distribution provides a good fit.
Following this work, we model the duration of each cluster as an independent gamma distribution. 
We use maximum likelihood estimation to fit the shape and rate parameters.

\subsection{Time-Stretching}
\label{sec:time-stretching}

To adjust rhythm, we up/down-sample the extracted soft units using linear interpolation.
By stretching the entire utterance, we can modify the overall speaking rate.
Alternatively, we can stretch individual segments for finer-grained control.

Based on global or fine-grained time-stretching, we propose two methods for rhythm conversion. 
For the global method, we first estimate the speaking rate of the source and target speakers.
Then we stretch the source utterance according to the ratio between the rates.

In the fine-grained approach, we stretch individual segments to match the target duration distributions.
Specifically, we use inverse transform sampling to map between the source and target distributions:
\begin{enumerate}
    \item For each cluster $c \in \{0,1,2\}$, we find the cumulative distribution functions for the source and target speakers: $F_{\text{src}, c}$ and $F_{\text{tgt}, c}$.
    \item Given a segment of source speech with duration $x>0$ belonging to cluster $c$, we compute $u = F_{\text{src}, c}(x)$.
    \item We find the corresponding target duration $y = F_{\text{tgt}, c}^{-1}(u)$ and stretch the segment according to the ratio $y/x$.
\end{enumerate}

\section{Experimental Setup}

Focusing on any-to-one conversion, we conduct three experiments to evaluate Urhythmic. 
The first experiment investigates the correlation between the syllable rate and estimated speaking rates. 
The second experiment compares the rhythm of the converted and target speech. 
The last experiment assesses naturalness, intelligibility and speaker similarity. 
We compare Urhythmic to three state-of-the-art unsupervised rhythm conversion systems: AutoPST~\cite{qian2021global}, UnsupSeg~\cite{kuhlmann2022investigation}, and DISSC~\cite{maimon2022speaking}.
We use the official pretrained models for each baseline.
Since UnsupSeg did not release the voice conversion component of their model, we apply their rhythm conversion method to unmodified outputs from Urhythmic.

We evaluate speaking rate estimation on LibriSpeech~\cite{panayotov2015librispeech}.
For the rhythm conversion experiment, we pick the three fastest and three slowest speakers from VCTK~\cite{veaux2017cstr}.
To avoid conflating accent and speaker identity, we limit the selection to a single region (Southern England).
For the subjective evaluations, we use the first 24 parallel utterances from each speaker.

For segmentation, we set the regularizer weight to $\gamma=2$.
We use HiFi-GAN~\cite{kong2020hifi} as the vocoder and adapt the generator to produce 16~kHz audio directly from soft speech units.
We pretrain the vocoder on LJSpeech~\cite{ito2017lj} for 3M steps. 
For each target speaker, we finetune a separate model for 50k steps.

\subsection{Speaking Rate Estimation}
\label{sec:exp-speaking-rate}

In the first experiment, we measure the correlation between the true syllable rate and estimated speaking rates.
Using forced alignments~\cite{mcauliffe2017montreal}, we calculate the average syllable rate for each speaker in the LibriSpeech dev and test split.
We remove utterances containing out-of-vocabulary words and filter silences from the alignments.

For DISSC, we estimate speaking rate by counting the number of deduplicated discrete units per second.
For AutoPST, we use the self-expressive autoencoder to define segment boundaries at points where the similarity between neighboring frames drops below a threshold.
We count the number of segments per second to estimate speaking rate.

\begin{table}[!b]
\mytable
    \caption{Correlation of estimated speaking rates with syllable rate.}
    \tablecaptionsep
    \begin{tabularx}{1.0\linewidth}{@{}lCCCc@{}}
        \toprule
         & Urhythmic & AutoPST & UnsupSeg & DISSC \\
        \midrule
        $r$ & $0.95 \pm 0.02$ & $0.33 \pm 0.15$ & $0.61 \pm 0.10$ & $0.82 \pm 0.05$ \\
        \bottomrule
    \end{tabularx}
    \label{tbl:speakingrate}
\end{table}

Table~\ref{tbl:speakingrate} reports the Pearson correlation coefficients $r$ with $95\%$ confidence intervals.
Urhythmic outperforms the baselines, showing a stronger correlation with the syllable rate.
This indicates that the speech segments discovered by Urhythmic allow for more accurate modeling of speaking rate.

\subsection{Rhythm Conversion}
\label{sec:exp-rhythm-conversion}

The second experiment compares the rhythm of the converted and target speech.
We align text transcriptions to the set of parallel utterances from VCTK and compute three metrics: phone length error (PLE), word length error (WLE), and total length error (TLE)~\cite{maimon2022speaking}.
These metrics measure duration differences at distinct scales, with lower error rates indicating a closer match to the target rhythm.

To avoid requiring parallel data for evaluation, we propose additional metrics comparing phone duration distributions between the converted and target speech. 
We use forced alignments to measure phone durations, group the data by sound type, and calculate the Wasserstein distance between the empirical distributions.
Smaller distances represent more similar duration distributions, implying better rhythm conversion.

Table~\ref{tbl:length-errors} reports TLE, WLE, and PLE for Urhythmic and the baseline systems.
As a reference, we also include voice converted speech from Urhythmic without rhythm modification.
Urhythmic improves all three metrics. 
Our global and fine-grained methods give comparable results at word and phone scales; however, fine-grained conversion substantially reduces TLE.
The improvement is explained by better silence modeling since TLE includes silences (unlike the other metrics).
We can clearly see this distinction in Table~\ref{tbl:wassertein}, which reports Wasserstein distances broken down by sound type.
While our global and fine-grained methods perform similarly across the different sound types, fine-grained modeling substantially improves the conversion of pauses and silences.

\begin{table}[!t]
    \mytable
    \caption{Total Length Error (TLE), Word Length Error (WLE), and Phone Length Error (PLE) on parallel utterances from VCTK.}
    \tablecaptionsep
    \begin{tabularx}{1.0\linewidth}{@{}llCCC@{}}
    \toprule
        Method & Scope & TLE & WLE & PLE \\
        \midrule
        AutoPST~\cite{qian2021global} & Fine & 6.94 & 0.233 & 0.073 \\
        UnsupSeg~\cite{kuhlmann2022investigation} & Global & 1.79 & 0.066 & 0.024 \\
        DISSC~\cite{maimon2022speaking} & Fine & 1.10 & 0.051 & 0.022 \\
        \addlinespace
        \multirow{2}{*}{Urhythmic} & Global & 1.01 & 0.046 & 0.020 \\
        & Fine & 0.78 & 0.047 & 0.020 \\
        \addlinespace
        \multicolumn{2}{@{}l}{No-Modification} & 1.56 & 0.058 & 0.022 \\
        \bottomrule
    \end{tabularx}
    \label{tbl:length-errors}
    \vspace*{-5pt}
\end{table}

\subsection{Naturalness, Intelligibility, and Speaker Similarity}
\label{sec:exp-subjective}

Next, we evaluate the naturalness, intelligibility, and speaker similarity of the converted speech.
To assess intelligibility, we measure word error rate (WER) using the Whisper-Small speech recognizer~\cite{radford2022robust}.
Lower WER indicates better intelligibility since the content of the source speech remains recognizable after conversion.

Following~\cite{das2020predictions}, we evaluate speaker similarity using a trained speaker-verification system.
An equal error rate (EER) of 50\% indicates high speaker similarity since the verification system cannot distinguish between converted speech and genuine examples from the target speaker.

Finally, we conduct subjective evaluations to assess naturalness and speaker similarity.
Using Prolific~\cite{palan2018prolific}, we recruited 84 English-speaking raters for each evaluation.
We followed the P.808~\cite{itu2018p808} recommendation and recorded 576 ratings (96 per method).
For naturalness, we report a mean opinion score~(MOS).
For speaker similarity, we follow the same/different protocol from the Voice
Conversion Challenges~\cite{wester2016analysis, yi2020voice}.
We pair each converted example with a reference from the target speaker and ask evaluators to rate their similarity on a four-point scale.
We aggregate the ratings into a mean similarity score (SIM).

Table~\ref{tbl:subjective} reports results for intelligibility (WER), naturalness (MOS), and speaker similarity (EER and SIM).
Urhythmic outperforms the baselines across all four metrics. 
Our global and fine-grained methods perform similarly with both improving subjective similarity scores compared to the no-modification reference.
We suspect this is because the evaluators account for differences in rhythm and prosody in their assessments of speaker similarity. 
Finally, Urhythmic achieves comparable WER and MOS to the no-modification reference, demonstrating that our approach to rhythm conversion has minimal impact on intelligibility and naturalness. 

\begin{table}
    \mytable
    \caption{Comparison between the duration distributions of the target and converted speech.}
    \tablecaptionsep
    \begin{tabularx}{1.0\linewidth}{@{}llCCCCCC@{}}
    \toprule
    Method & Scope & vowel & appx. & nasal & fric. & stop & sil. \\
    \midrule
    AutoPST~\cite{qian2021global} & Fine & 110 & 18 & 317 & 32 & 138 & 3647 \\
    UnsupSeg~\cite{kuhlmann2022investigation} & Global & 16 & 10 & 14 & 16 & 15 & 294 \\
    DISSC~\cite{maimon2022speaking} & Fine & 7.6 & 4.4 & 6.6 & 7.1 & 9.0 & 208 \\
    \addlinespace
    \multirow{2}{*}{Urhythmic} & Global & 5.9 & 6.5 & 4.6 & 7.1 & 6.1 & 207 \\
    & Fine & 6.3 & 5.5 & 5.6 & 7.2 & 5.3 & 70 \\
    \addlinespace
    \multicolumn{2}{@{}l}{No-Modification} & 13 & 8.2 & 11 & 13 & 12 & 270 \\
    \bottomrule
    \end{tabularx}
    \label{tbl:wassertein}
    \vspace*{-5pt}
\end{table}

\begin{table}[!t]
    \mytable
        \caption{Voice conversion results for intelligibility (WER), speaker similarity (EER, SIM) and naturalness (MOS).
        }
    \tablecaptionsep
    \begin{tabularx}{1.0\linewidth}{@{}llCCcc@{}}
        \toprule
        Method & Scope & WER & EER & MOS & SIM \\
        \midrule
        AutoPST~\cite{qian2021global} & Fine & 80.3 & 9.7 & $1.34 \pm 0.08$ & $2.10 \pm 0.15$ \\
        DISSC~\cite{maimon2022speaking} & Fine & 17.5 & 40.0 & $2.24 \pm 0.11$ & $2.16 \pm 0.17$ \\
        \addlinespace
        \multirow{2}{*}{Urhythmic} & Global & 3.3 & 46.0 & $3.67 \pm 0.12$ & $3.56 \pm 0.10$ \\
        & Fine & 3.4 & 47.3 & $3.74 \pm 0.14$ & $3.41 \pm 0.12$ \\
        \addlinespace
        \multicolumn{2}{@{}l}{No-Modification} & 3.3 & 47.4 & $3.85 \pm 0.12$ & $3.25 \pm 0.12$ \\
        \multicolumn{2}{@{}l}{Ground Truth} & 2.7 & - & $4.02 \pm 0.14$ & $3.63 \pm 0.13$ \\
        \bottomrule
    \end{tabularx}
    \label{tbl:subjective}
    \vspace*{-2pt}
\end{table}

\section{Conclusion}

We proposed Urhythmic, an unsupervised approach to rhythm and voice conversion.
We presented methods for modeling global and fine-grained characteristics of rhythm. 
The global method estimates overall speaking rate, while the fine-grained method models the duration distribution of discovered speech units.
Results show that the estimated speaking rate correlates well with the syllable rate, and that fine-grained conversion accurately models the target speaker's rhythm.
Finally, Urhythmic outperforms other unsupervised rhythm conversion systems in subjective and objective evaluations.

\vfill\pagebreak

\bibliographystyle{IEEEtran}
\bibliography{IEEEabrv, references}

\end{document}